\documentclass[final,5p,authoryear,twocolumn]{elsarticle}

\usepackage{amssymb}
 \usepackage{lineno}

\usepackage{url}
\usepackage{hyperref}
\hypersetup{
    colorlinks=true,
    citecolor=blue,
    linkcolor=red,
    filecolor=magenta,      
    urlcolor=cyan,
}
\usepackage{amsmath}
\usepackage{booktabs} % For pretty tables
\usepackage{caption} % For caption spacing
\usepackage{subcaption} % For sub-figures
\usepackage{graphicx}
\usepackage{pgfplots}
\usepackage[all]{nowidow}
\usepackage[utf8]{inputenc}
\usepackage{tikz}
\usetikzlibrary{er,positioning,bayesnet}
\usepackage{multicol}
\usepackage{algpseudocode,algorithm,algorithmicx}
\usepackage{listings}

\usepackage{tikz}
\usepackage{expl3}
\usepackage{amsmath, amssymb}
\usepackage{xcolor}

\usetikzlibrary{positioning,calc}
\usetikzlibrary {shapes,matrix}
\usetikzlibrary{arrows}

\usepackage[inline]{enumitem} % Horizontal lists
\definecolor{blue}{HTML}{1F77B4}
\definecolor{orange}{HTML}{FF7F0E}
\definecolor{green}{HTML}{2CA02C}

\pgfplotsset{compat=1.14}

\begin{document}

\begin{frontmatter}
\title{Large-scale Foundation Models and Generative AI for BigData Neuroscience}
\author[inst1]{Ran Wang}
\affiliation[inst1]{organization={Department of Psychiatry, New York University Grossman School of Medicine},
            city={New York},
            state={NY},
            postcode={10016}, 
            country={USA}}

\author[inst1,inst2,inst3]{Zhe Sage Chen}

\affiliation[inst2]{organization={Department of Neuroscience and Physiology, Neuroscience Institute, New York University Grossman School of Medicine},%Department and Organization
            city={New York},
            state={NY},
            postcode={10016}, 
            country={USA}}

\affiliation[inst3]{organization={Department of Biomedical Engineering, New York University Tandon School of Engineering},%Department and Organization
            city={Brooklyn},
            state={NY},
            postcode={11201}, 
            country={USA}}

%
%\maketitle              % typeset the header of the contribution
%
\begin{abstract}
Recent advances in machine learning have made revolutionary breakthroughs in computer games, image and natural language understanding, and scientific discovery. Foundation models and large-scale language models (LLMs) have recently achieved human-like intelligence thanks to BigData. With the help of self-supervised learning (SSL) and transfer learning, these models
may potentially reshape the landscapes of neuroscience research and make a significant impact on the future.
Here we present a mini-review on recent advances in foundation models and generative AI models as well as their applications in neuroscience,  including natural language and speech, semantic memory, brain-machine interfaces (BMIs), and data augmentation. We argue that this paradigm-shift framework will open new avenues for many neuroscience research directions and discuss the accompanying challenges and opportunities.

\end{abstract}



\end{frontmatter}

\section{Introduction}

Advances in neurotechnology have allowed us to record large-scale, high-throughput neural data through in vivo electrophysiology and brain imaging.  These BigData present a challenge for various neural data analyses such as decoding and functional connectivity analysis, as well as closed-loop brain-machine interface (BMI) applications in neuroscience experiments \citep{ChenPesaran21}. In parallel, machine learning research is also moving very fast. Rapid advances in deep learning and development of large-scale foundation models and large language models (LLMs) have taken the whole world by storm, demonstrating remarkable and revolutionary findings in generating high-resolution synthetic images, yielding human-like natural language understanding and human-level creativity \citep{zhao2023survey,naveed2023comprehensive,Singhal23Nature}. Without exaggeration, the past few years have witnessed a paradigm shift in AI to foundation models in nearly every aspect of machine learning applications. {\it How will these technological changes impact and imply for neuroscience?} Answers to this question are part of our motivations to write this review. However, since the field is relatively new, the number of published studies on neuroscience applications based on foundation models or LLMs is relatively small, but the interest is rapidly growing and many findings  derived from this line of research may have a potentially significant impact  on neuroscience.

In this mini-review, we first provide a brief overview of foundation models, its building block---transformers, and extend our overview to a broad class of generative AI tools. Further, we will review important concepts in representation learning, self-supervised learning (SSL) and transfer learning, which will play important roles in cross-modality applications. Next, we will review recent applications of foundation models and generative AI in various neuroscience research areas, including but not limited to large-scale brain imaging data analysis, natural speech and language understanding, memory, emotion, mental state decoding,  behavior, BMI, and data augmentation.  Finally, we conclude the review with discussions and outlook on future research opportunities and challenges.

\section{Foundation models and generative AI }

\subsection{What are foundation models?}

\begin{figure}
\centering
\includegraphics[width=\columnwidth]{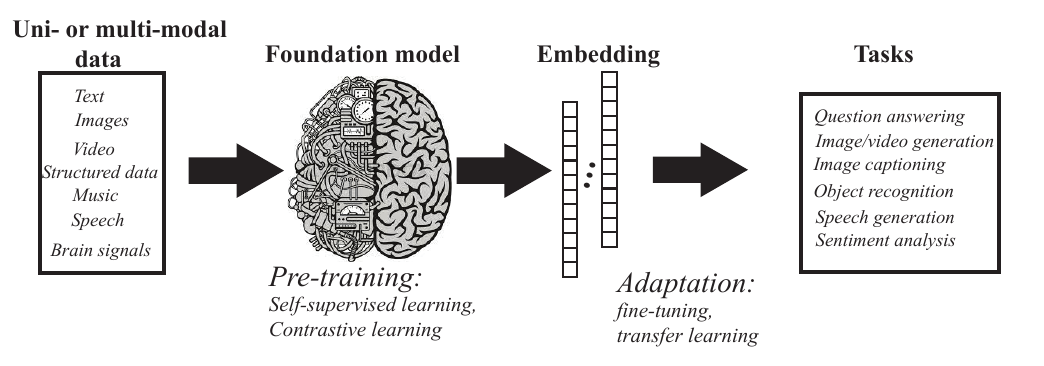}
\caption{A schematic diagram of foundation models.}
\label{Fig1}
\end{figure}

A foundation model is a ``paradigm for building AI systems" in which a model trained on a large amount of unlabeled data can be adapted to many other applications. The foundation models 
 are often trained using self-supervision with BigData, and can be adapted to a wide range of tasks (e.g., text, images, speech, structured data, brain signals, and high-dimensional tensor data) (Fig.~\ref{Fig1}).
 One of the popular class foundation models is LLMs (Table~\ref{LLMs}), which take language input and generate synthesized output. In general, foundation models work with multi-modal data types. 

 In a recent group study conducted at Stanford university, it was concluded that ``{\it foundation models are  scientifically interesting due to their impressive performance and capabilities, abut what makes them critical to study is the fact that they are quickly being integrated into real-world deployments of AI systems with far-reaching consequences on people}'' \citep{Bommasani2021FoundationModels}.

 At the very high level, there are two fundamental ideas in the LLM and foundation models: (i) embedding, which aims to convert words or tokens into high-dimensional statistically meaningful numbers; (ii) SSL or contrastive learning. 

 \begin{table*}
\centering
    \caption{A selective list of foundation models and LLMs.} 
 %   \resizebox{\columnwidth}{!}{
  \begin{tabular}{lll}
  \hline

\bf Model  &	\bf Characterization &	\bf Developer \\ \hline

 BERT   &generative language model& 	Google  \\ 

 CLIP   &language-image pre-training & Open AI	  \\ 

  Codex   &  general-purpose programming model & 	Open AI  \\ 

DALE-E, DALL-E2   & text-to-image models & 	Open AI  \\ 

% Claude   & .... model & 	Anthropic  \\ 

GPT-3   & causal sequence model for NLP & Open AI	  \\ 

GPT-4   & multi-modal model & Open AI	  \\

PaLM, PaLM2  & multi-lingual pathways language models & Google  \\ 

LLaMA, LLaMA2   & foundational language model, code generation model & 	Meta  \\ 

SEER & self-supervised computer vision model& 	Meta  \\

GATO  & multi-modal, multi-task, multi-embodiment policy & DeepMind	  \\ % sequence-to-sequence model

DINOv2 & foundational models for vision & Meta\\ 

  \hline
\end{tabular}
%}
\label{LLMs}
\end{table*}

\subsubsection{Embedding}

Embedding is a feature extraction technique that nonlinearly transforms the input signal to a representational vector that are easy to indexed, searched, computed, and visualized. 
In language processing applications, a word embedding is to project words onto a meaningful space in which words ``are nearby in meaning'' appear nearby in the embedding. Take ChatGPT as an example, the dimensionality of the embedding space can be high-dimensional (hundreds to thousands depending on the specific layer). Therefore, the embedding vectors that contains a string of numbers 
are located in the coordinates of ``linguistic feature space''. 
In deep neural networks, embedding layers enable us to  learn the relationship between high-dimensional inputs and outputs more efficiently.

 \subsubsection{SSL learning}

In real life, humans and animals can learn efficiently from observation or very few labeled examples, pointing the limitation of BigData-based supervised learning. SSL is predictive learning in that it aims to 
predict missing parts of the input. In recent years, SSL techniques have achieved immense successes in  natural language processing (NLP) and computer vision by enabling models to learn from BigData at unprecedented scales \citep{millet2023realistic,balestriero2023cookbook}.  Depending on the objective, SSL can be a generative, contrastive, or generative-contrastive (adversarial) form; a comprehensive survey of SSL is referred to elsewhere \citep{Liu23}.   Under the SSL framework, fine-tuning the pre-trained models with a small percentage of labeled data can achieve comparable results with the supervised training \citep{emadeldeen2023eval}. In NLP, pre-training methods like BERT (Bidirectional Encoder Representations from Transformers)  have shown strong performance gains using SSL that masks individual words or subword units \citep{BERT}. Recently, \cite{Joshi20} proposed an extended version of BERT  known as SpanBERT, which can mask continguous random spans instead of random tokens and train the span boundary representations to better predict the entire content of the masked span; by so doing, SpanBERT consistently outperforms BERT, with the largest gains on span selection tasks.

\subsection{Transformer model }

\begin{figure*}[t]
\centering
\includegraphics[width=0.9\linewidth]{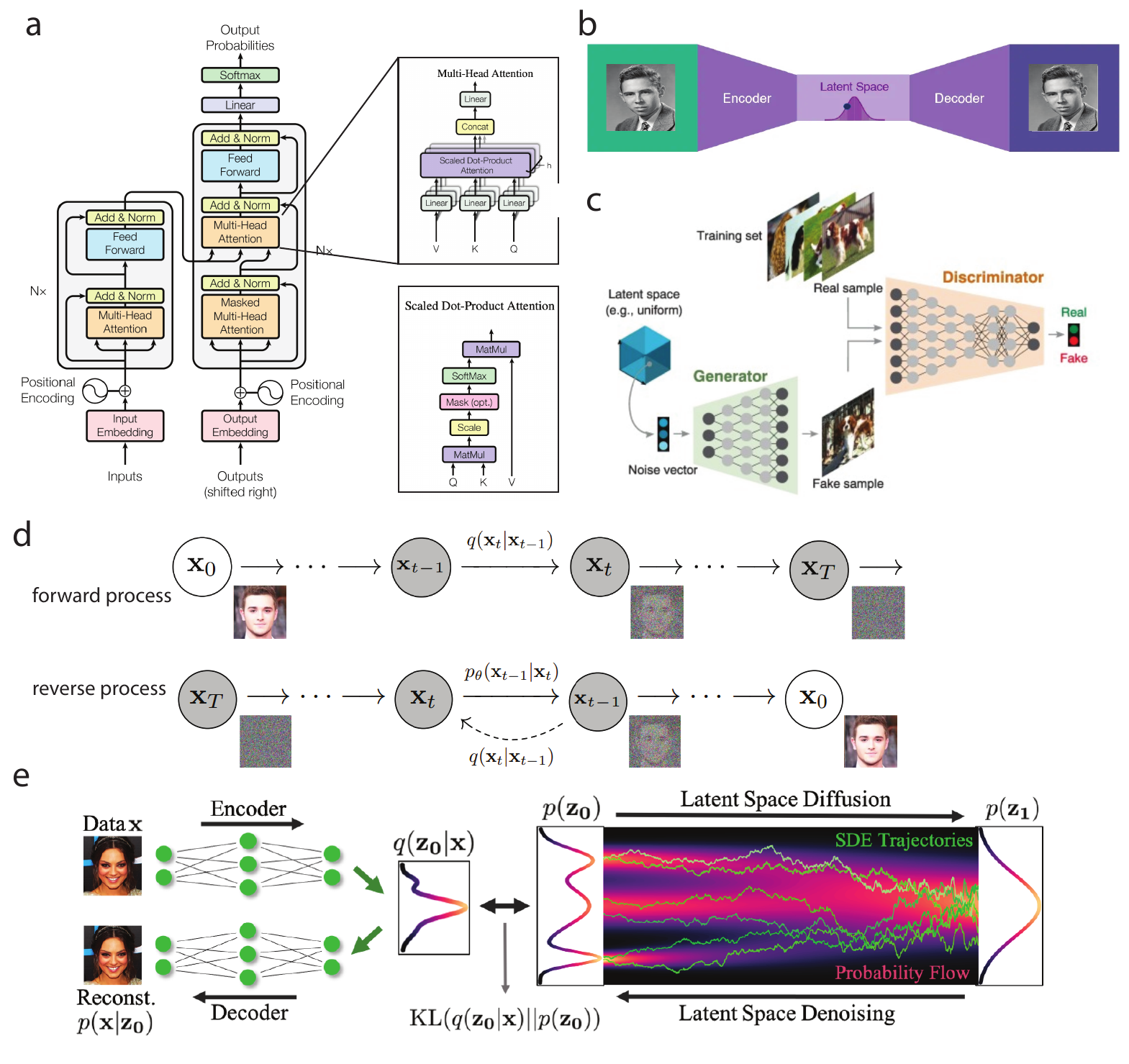}
\caption{Schematics of several generative AI models: (a) the transformer architecture, (b) VAE, (c) GAN, (d) diffusion model. (e) LSGM. Panel a is adapted with permission from \citep{vaswani2017attention}. Panel c is adapted with permission from \citep{GAN21}, Elsevier. Panel d is adapted with permission from \citep{Ho20}. Panel e is adapted with permission from \citep{vahdat2021NIPS}. }
\label{Fig2}
\end{figure*}

A transformer model is a deep neural network that learns context and thus meaning by tracking relationships in sequential data.
Specifically, transformers were developed to solve the problem of sequence transduction that transforms an input sequence to an output sequence, enabling end-to-end learning in machine translation, text generation and sentiment analysis \citep{vaswani2017attention}. Transformers are the building blocks in many foundation models, such as 
BERT  and GPT (Generative Pre-trained Transformer). Transformers  are computationally efficient in simultaneous sequence processing since model training can be sped up through parallelization, a key feature missing in recurrent neural networks (RNNs) and long short-term memory (LSTM); this feature has also made the creation of LLMs feasible.

The transformer model has a seq2seq  neural network architecture, consisting of encoding, decoding and self-attention modules (Fig.~\ref{Fig2}a). There are several concepts fundamental to computations in the transformer:
\begin{itemize}
\item word embeddings: vector representations of words
\item positional embeddings: encoding the position of each token in a sequence and add the positional information to the word embeddings 
\item attention: understanding the context of a word by considering words that go before or after it. In other words, if the meaning is a result of relationships between things, then self-attention is a general way of learning relationships \citep{vaswani2017attention}.
\item self-attention:  weighing the importance of different parts of the input sequence against each other. 
\item multi-head attention: allowing the network to learn multiple ways of weighing the input sequence against itself.
\end{itemize}

In addition to NLP applications, the transformer architecture has been applied in other domains such as computer vision \citep{dosovitskiy2020image},  visual stimulus classification \citep{Bagchi22}, neural data analysis \citep{ye2021representation}, and reinforcement learning (RL) \citep{li2023survey}.

\subsection{Generative AI}

Generative AI describes a class of algorithms that can be used to create new content, including audio, code, images, text, simulations, and videos. Several representative generative AI algorithms are summarized below. 

\begin{itemize}

\item Variational Autoencoder (VAE): VAE is a generative AI algorithm that uses deep learning to generate new content, detect anomalies and remove noise \citep{kingmaVAE}. VAE consists of an encoder and a decoder, separated by the latent space (Fig.~\ref{Fig2}b).  The latent space contains an abstract representation of the data containing only the most meaningful information (i.e., dimensionality reduction). The model can learn the data distribution, so that a corresponding output can be reconstructed based on a new sample input.

\item Generative Adversarial Network (GAN): A GAN is a class of deep learning framework that uses two neural networks, the generator and the discriminator (Fig.~\ref{Fig2}c), to generate new and realistic synthetic data that are similar to the samples among the training set. Specifically, the generator network takes random noise as input and generates synthetic data, and is aimed  to produce data that are indistinguishable from the real data in the training set. The generator tries to create realistic samples  and follow the patterns present in the original dataset.
On the other hand,  the discriminator network evaluates the data it receives and tries to distinguish between real data from the training set and the synthetic data produced by the generator. Its goal is to correctly classify whether the input data is real or generated by the generator. The discriminator provides feedback to the generator, helping it improve its generated samples. To date, the GAN and many of its variants have numerous applications in image generation, image-to-image translation, super-resolution imaging, text-to-image synthesis, and video generation \citep{GAN14,gui2020review,GAN21}.

\item Generative Pre-trained Transformer (GPT): GPT is specifically referred to  a series of language models that use the transformer architecture to understand and generate coherent and contextually relevant text. Because of powerful predictive ability, GPT is effective for a variety of NLP tasks, including text generation, translation, and summarization.
The basic idea behind GPT is to apply SSL and train 
large datasets containing a diverse range of text from various sources. Upon the completion of learning, the model 
takes  the sequence of tokens that corresponds to the text in the past and finds an embedding that represents them, and further generate a large number of values that turn into probabilities for predicting possible next tokens \citep{Wolfram23}. The newer GPT developments, such as GPT-3 \citep{GPT20} and GPT-4, represent a landmark in this technology.

\item Diffusion Model: Diffusion models 
are referred to a class of latent generative models that are used to model the distribution of data based on Markov chains and variational inference (Fig.~\ref{Fig2}d) \citep{Ho20,rombach2021highresolution}. These models are designed to capture the underlying data distribution by iteratively transforming a simple distribution into a complex one. Diffusion models offer a promising avenue for deep generative modeling owing to  robust expressive capacity, and ability to generate data via ancestral sampling without the prerequisite of a posterior distribution. 
Unlike other deep generative models such as VAE and GAN, training diffusion models is relatively simple. 
 To date, diffusion models have been used in image generation, NLP, and time series analysis.

 \item Latent Score-based Generative Model (LSGM): The LSGM generalizes the ideas of VAE and diffusion model, maps the input onto a latent space and applies the diffusion model in the latent embeddings of the data (Fig.~\ref{Fig2}e) \citep{vahdat2021NIPS}. As an extension to score-based generative models \citep{Song19NIPS,song2021scorebased}, the LSGM has several key computational advantages: synthesis speed, expressivity, and tailored encoders and decoders.

\end{itemize}

Foundation models can serve as a basis  for generative AI. BERT and GPT models have already been used as 
the building blocks for developing more sophisticated generative AI models. For instance,  
 \cite{fei2022towards} developed a self-supervised pre-trained foundation model on vision-language multi-modal input, which only requires weak semantic correlated image-text training pairs; specifically, they demonstrated that the foundation model not only can generate high-level concepts and describe complicated scenes, but also has an ability to imagine, which represents a step towards artificial general intelligence (AGI). 

 Furthermore, foundation models may provide a starting point for developing more advanced generative AI systems. Researchers and developers often fine-tune or extend the foundation models to create specialized generative models tailored to specific tasks or domains. More importantly, foundation models may facilitate transfer learning, which is vital for generative AI--as it allows models to leverage the knowledge and representations learned by foundation models to generate diverse and contextually appropriate content across different domains. 
One  exciting application of generative AI is to decode brain signals  and transform them into text or images, which may have a translational impact on the lives of individuals with  traumatic brain injury (TBI) or server paralysis who cannot communicate through speech, typing, or gestures \citep{Metzger23,Metzger22,defossez2022decoding,VanRullen19_CB,tang2023semantic}. Recently, GAN-based  \citep{Dado22} and diffusion model-based \citep{takagi2023high} approaches have  been developed to reconstruct human faces or visual images from fMRI recordings.
See \citep{Gong23} for a short review on generative AI for brain imaging applications.

\section{Representation learning and transfer learning}

\subsection{Representation learning}

Representation Learning is referred to a class of machine learning algorithms that extract meaningful patterns from raw data to create representations easily understood or processed  \citep{bengio2014representation}. During this process, dimensionality reduction, regularization,  invariance, and sparsity play an important role. Current LLMs heavily rely on effective representation learning algorithms. Representation learning can be achieved  by unsupervised, supervised, and self-supervised frameworks. For instance, as a special case of SSL paradigm, contrastive learning can  learn an embedding space such that similar instances have close representations while dissimilar instances stay far apart from each other.  In addition to computer vision and NLP tasks, contrastive learning has  been used to extract meaningful  representations from neural data, including data from electroencephalography (EEG), magnetoencephalography (MEG), functional magnetic resonance imaging (fMRI), and other neuroscience modalities \citep{Kostas21}. For instance,  contrastive learning has enabled researchers to uncover patterns in brain connectivity data, providing insight into the organization and communication between different brain regions, or identifying connectivity-based biomarkers between healthy and pathological brains \citep{Tong23}. Contrastive learning can also learn representations in the latent feature space based on dimensionality reduction. One such an example  is contrastive PCA (cPCA), which can identify the dominant subspace that distinguishes two datasets collected from different conditions \citep{Abid18}.  Additionally, contrastive variational autoencoder (cVAE) \citep{Aglinskas22}, as an extension to cPCA, offers a more flexible approach capable of modeling nonlinear relationships between the inputs and latent features. Finally, another contrastive learning paradigm, 
contrastive predictive coding (CPC) \citep{oord2019representation}, learns self-supervised representations by predicting the future in latent space by using autoregressive  models and VAE; the model uses a probabilistic contrastive loss which induces the latent space to capture information that is maximally useful to predict future data.

\subsection{Transfer learning}

Transfer learning represents a class of machine learning technique where knowledge learned from a task is reused in order to boost performance on a related task or generalize out-of-distribution via targeted re-training \citep{Pan10}. In deep learning models, transfer learning has been widely used in computer vision, image classification, and NLP tasks  \citep{Yosinski14,Goodfellow16}.

Transfer learning has found many applications in neuroscience. 
In neuroimaging data analysis, pre-trained models from NLP or computer vision domains, can be fine-tuned or used to extract features from raw neural data, facilitating out-of-domain tasks such as classification, segmentation, and decoding of neural activity. For instance, pre-trained models from related medical imaging tasks can be adapted to process and interpret neuroimaging data, leading to a more accurate and efficient analysis. Additionally, since the relationship between cognitive tasks is usually represented by similarity of neural representations or activated brain regions, transfer learning may perform better in task decoding with fMRI data if the source and the target cognitive tasks activate similar brain regions \citep{qu2022transfer}. 

In BMI research, transfer learning can improve the performance and adaptability of BMI systems by leveraging knowledge from related tasks. Pre-trained models may help enhance the decoding of neural signals for controlling external devices or for interpreting brain activity associated with specific motor or cognitive tasks.

Transfer learning can assist in the early detection and diagnosis of neurological or psychiatric disorders by leveraging knowledge from related medical domains. Pre-trained models from medical imaging or clinical data analysis can be adapted to identify biomarkers  associated with specific pathological conditions, aiding in early intervention and personalized treatment strategies. Notably, transfer learning can work well where the data sample size is small in neuroimaging-based prediction \citep{Ardalan22,Malik22} and ECoG/EEG decoding analysis \citep{Zhang2021,Peterson21}.

\section{Foundation models and generative AI  for neuroscience applications}

\subsection{ Context-dependent embedding mapping }

As discussed earlier, representation learning can identify context-depending embeddings for a broad class of input signals.
For instance, if the input is a speech signal, the embedding mapping for speech representation may be produced by 
 ``wave2vec'' \citep{schneider2019wav2vec,baevski2020wav2vec,millet2023realistic}, HuBERT \citep{hsu2021hubert}, and ``data2vec'' \citep{baevski2022data2vec,baevski2023efficient}.
If the input is a neural time series such as EEG signal, 
the embedding mapping for EEG may include ``EEG2vec''  \citep{EEG2vec23} or other representation learning methods 
 \citep{Kostas21,Rafiei22SSL,pmlr-v158-wagh21a}. Such methods have been demonstrated in neuroscience applications such as   automatic sleep staging \citep{banville2020uncovering,Yang23SSL} and seizure detection \citep{Tang22SSL}.

In neural data analysis, embeddings have been widely adopted in unsupervised or supervised representation learning. For instance, automated neuron reconstruction and annotation of volume electron microscopy (VEM) datasets of three-dimensional images of brain tissue is computationally intensive and challenging. \cite{Schubert19} first used unsupervised training to infer morphology embeddings (``neuron2vec'') of neuron reconstructions, and then trained cellular morphology neural networks (CMNs) to identify glia cells via supervised classification; they also demonstrated in using CMNs to identify subcellular compartments and the cell types of neuron reconstructions.

Embeddings are useful for revealing low-dimensional neural dynamics and modeling naturalistic behaviors \citep{Wang22,Schneider23}. Although traditional latent variable models  have been used for analyzing neural and behavioral data \citep{Chen15,Latimer15,Calhoun19,Bolkan22,Ashwood22,Lakshminarasimhan23}, most of them are limited in encoding the context dependence. Incorporating task-relevant embedding vectors to form a context-relevant embedding would allow us to perform end-to-end learning efficiently. Recently, \cite{ye2021representation} have proposed a non-recurrent, BERT encoder-based neural data transformer (NDT) model  to explicitly model autonomous neural population activity and reported comparable performance between the NDT model and other RNN models. In their NDT model, inputs to transformer layers were first normalized and enriched through contextual information
(``self-attention'' blocks), and passed through a feedforward module.

\subsection{Brain imaging}

Human neuroimaging provides a window to examine a healthy and diseased brain, in terms of both structural and functional forms, including EEG, MEG, fMRI, diffusion tensor imaging (DTI),  and positron emission tomography (PET).
See \citep{Gong23} for a review of generative AI for brain imaging, covering 
co-registration, super-resolution, enhancement, classification, segmentation, cross-modality, brain network analysis, and decoding analysis.

 Several lines of work have proposed generative AI approaches to reconstruct visual images based on fMRI data \citep{Seeliger18,VanRullen19_CB,ferrante2023semantic}. For instance, \cite{VanRullen19_CB} first trained a VAE  network using a GAN  unsupervised procedure over a large dataset of celebrity faces, where the VAE latent space provided a topologically organized 1024-dimensional embedding of each image. Next, they  presented  thousands of face images to human subjects, and learned a linear mapping between multi-voxel fMRI activation patterns and latent embeddings. Finally, they applied this mapping to novel face images, translating fMRI patterns into reconstructed faces.

 \cite{lu2022multimodal} developed a self-supervised pre-trained image-text multi-modal foundation model which outperformed CLIP (Contrastive Language-Image Pre-Training) model even with a small percentage ($\sim$3.75\%) of training pairs. The image and text were first encoded individually by pre-trained uni-modal large-scale models, vision transformer (ViT) and BERT. The output of BERT was then projected to  a trained mapping layer that aligns with ViT features. By comparing the encoded image encoding feature with fMRI imaging of the human visual cortex, their results showed that the proposed multi-modal model has higher prediction accuracy than the uni-modal image encoder.

\subsection{Natural language and speech}

Speech and language understanding involves 
a deep comprehension of their generation and processing (in both sound and text), enabling computers to perform tasks such as speech recognition, language translation, sentiment analysis, and text summarization.

Representing human speech from brain signals (such as ECoG and fMRI) consists in decoding neural activity  associated with speech production, perception, or comprehension.  
It has been known that natural speech reveals a semantic map that tiles the human cerebral cortex 
\citep{Hux16Nature}, and the semantic space is  continuously distributed across the brain describing representations of thousands of object and action categories \citep{Hux12,Hux16}.

On the one hand,  the rich features extracted from the  foundation models provide a new hypothesis when studying brain representations during specific speech and language tasks. For example, the ECoG activity in the superior temporal gyrus (STG) and inferior frontal gyrus (IFG) of the human brain was found to be correlated with features extracted by the GPT model \citep{goldstein2022shared}. Since predictive pre-training of the GPT model was capable of encoding contextual information, word onset, and word surprisal, this finding suggests that the human auditory cortex may encode speech in a similar manner. The contextual encoding phenomenon was also found when correlating neural representations in the human auditory cortex with the HuBERT model's embeddings \citep{li2022dissecting}.

On the other hand, a growing number of studies have focused on decoding human speech  from invasive brain recordings, using either intracranial ECoG or intracortical spiking activity \citep{Metzger22,Moses21,willett2023high} (see the review of BMI applications below). 
Recently, \cite{defossez2022decoding} have developed a contrastive learning approach to decode speech based on  non-invasive magneto- or electro-encephalography (MEG/EEG). They first employed a large-scale pre-trained speech encoding model (``wave2vec 2.0'' \citep{baevski2020wav2vec}) to extract semantic features from speech, and then trained a decoding model to extract  features that converged to the speech features of corresponding trail while diverging from speech features of other trails. The model was capable of identifying the speech segment with features that best matched the decoded neural features. This work represents a large step forward in clinical practice without putting patients at the risk of brain surgery. 

Furthermore, EEG signals can be leveraged 
 to augment multi-modal NLP models while using less training data
\citep{Hollenstein21}; in combination with EEG data, BERT embeddings have showed consistently improved performance for NLP tasks.

\subsection{Memory and semantic reconstruction}

In the traditional episodic memory paradigm, subjects are usually required to memorize arbitrary items (words or images), lacking the fundamental components in real-life naturalistic events occurring over a longer timescale. Multimedia stimuli such as music and film, however, may provide rich contextual and naturalistic memory behaviors \citep{Groussard09}.
%In film, semantics deals with words, meanings (building blocks for genre), costume, acting, cinematography, etc., and syntax deals with how sentences are constructed to  create a deeper meaning of the film. 
% 8 narrative elements of a movie: theme, plot, characters, setting, conflict, point-of-view, tone and style.
%Music and film has defined ``event boundaries", which separate continuous sensory experiences into discrete meaningful events that can be  parsed by the brain \citep{Kurby08,Shin21}.

%The interplay of episodic and semantic memory in guiding repeated search in scenes

In neuroscience experiments, recollection of short audiovisual segments from movies can be viewed as a proxy to real-life memory that  consists of a stream of continuous sensory experiences. In contrast to pure reconstruction of static images from brain imaging \citep{Shen19,Horikawa13}, reconstructing high-quality images with correct semantics from brain recordings is more challenging  due to the complex underlying representations of brain signals and the scarcity of data annotations. In the literature, neural decoders have been developed for semantic reconstruction of movie or visual experiences \citep{Hux16,Nishimoto11}.  Extension of this framework using generative AI would represent a promising research direction.

 Recently, \cite{chen2023seeing} proposed a conditional diffusion model with sparse masked modeling for human visual decoding. Inspired by sparse coding in the primary visual cortex, they first  applied SSL and mask modeling in a large latent space for fMRI data; then they augmented a latent diffusion model (LDM) to reconstruct highly plausible images with semantically matching details from fMRI recordings using very few paired annotations.

\subsection{Mental state and emotion}

Decoding brain states and mental processes based on  brain imaging data has been an active research area \citep{Poldrack12,Rubin17}. However, the common challenge is that the sample size is relatively small and the  model is prone to overfitting. Recently, to decode mental states,   \cite{thomas2022selfsupervised}  proposed to leverage publicly shared fMRI data (\url{https://openneuro.org/}) to pretrain a foundation model. Their procedure consisted of two steps. In the first step, performing self-supervised learning on fMRI time series using various model strategies: seq-to-seq autoencoder, casual sequence modeling (similar to GPT-3), sequence-BERT, and network-BERT. In the second step, applying a plug-in and adaptation to decoding mental states. In so doing, the mental states can be viewed as a high-dimensional neural embedding, and NLP-inspired architectures were able to learn useful representations of fMRI time series; more importantly, the pre-trained model also improved the decoding accuracy of mental states (compared to several baseline models).

Decoding emotions from brain activity is one  fundamental task in human-computer interaction, yet most decoding methods are limited by the number of emotion category or has ignored the discrepancy of emotion expression between two brain hemispheres. Recently, \cite{Fu22} proposed a multi-view multi-label hybrid model for fine-grained emotion decoding: the generative component is a multi-view VAE that learns the brain activity of left and right hemispheres, as well as their differences; the discriminative component is a multi-label classification network; furthermore, they used a label-aware module for emotion-specific neural representation learning and modeled the dependency of emotional states by a masked self-attention mechanisms.

\subsection{Naturalistic behavior}

An important goal in neuroscience is to uncover the circuit mechanisms underlying cognitive processes and behavior, for which quantitative behavioral descriptions may play a vital role in linking brain activity and behavior \citep{Krakauer17,Pereira20}. Unlike constrained behaviors (such as head-fixed tasks or planar reach-and-grasp movement),  naturalistic behavior is  referred to the  behavior that animals have a tendency to exhibit under natural or realistic conditions, which is  often pleasurable and beneficial to biological functioning.

Given the success of sequence modeling in NLP, it is tempting to frame  behavior analysis as a sequence modeling problem and apply this idea to context-relevant behavioral embedding and attention computation. 
Recently, \cite{reed2022} have proposed a generalist agent (GATO) model for multi-modal, multi-task learning. Specifically, they encoded various modalities into a single vector space of ``tokens" that can be ingested by a large sequence model such as transformers; they also proposed various ``tokenization" approaches to capture the large amount of multi-modal data that include standard vision and language datasets and  some RL benchmarks.

\begin{table*}
\centering
    \caption{A representative list of recent BMI and neural decoding studies based on generative AI. } 
 %   \resizebox{\columnwidth}{!}{
  \begin{tabular}{llll}
  \hline

\bf Study  &	\bf Data & \bf Model &	\bf Application \\ \hline

\citep{anumanchipalli2019speech} & ECoG  & bidirectional LSTM &  brain2speech  \\  

\citep{wang2020stimulus} & ECoG & GAN, transfer learning & brain2speech \\ 

\citep{wang2023distributed} & ECoG & ResNet & brain2speech \\ 

\citep{Willett21}  &  ECoG & RNN, language model &  brain2text   \\  

\citep{willett2023high}  & microelectrode arrays  & RNN & brain2speech2text   \\ 

\citep{Metzger22} &  ECoG  & neural network & brain2text   \\ 

\citep{Metzger23} &  ECoG  & HuBERT, bidirectional RNN & brain2speech   \\ 

\citep{Liu23} &  ECoG  & sequential CNN-LSTM & brain2speech   \\

\citep{tang2023semantic}  & fMRI  & GPT-2 &  brain2text \\ 

\citep{takagi2023high} &  fMRI & diffusion model &  brain2image  \\ 

\citep{VanRullen19_CB} &  fMRI & VAE, GAN & brain2face   \\ 

\citep{ferrante2023semantic} & fMRI & CNN & brain2feature \\ 

\citep{ferrante2023brain} & fMRI &  generative image-to-text transformer &  brain2image\&text\\

\citep{huang2021deep} & fMRI &  deep VAE, LSTM &  brain2image\\

\citep{ozcelik2023natural} & fMRI & very deep VAE, diffusion model & brain2image \\

\citep{defossez2022decoding} & MEG, EEG & Contrastive Language-Image Pre-Training & brain2speech \\

\citep{Meta23} & MEG & DINOv2 & brain2image \\ 

\citep{bellier2023music} & ECoG & feedforward neural network  & brain2music  \\

\citep{denk2023brain2music} & fMRI & MusicLM  & brain2music \\

\citep{azabou2023unified} & spikes & PerceiverIO & brain2behavior \\

  \hline
\end{tabular}
%}
\label{BMI}
\end{table*}

\subsection{Brain-machine interfaces}

A BMI is a system that establishes a direct communication pathway between the brain's electrical activity and an external device, reading out the encoded stimuli (e.g., speech, vision, location) or translating thought into action (i.e., neuroprosthetics) \citep{Gilja12,Lebedev17,Willett21}. Such mind-reading devices can be used not only for translational applications \citep{Shanechi19,Moses21,Zhang21,Sun22}, but also for  scientific inquiry in basic science questions \citep{Sadtler14}.

Data sources in different BMIs have a varying degree of signal-to-noise ratio (SNR). For instance, while sharing the same temporal resolution, ECoG has a higher SNR than the scalp EEG. 
On the other hand, calcium imaging or fMRI data have a much lower temporal resolution than ECoG or EEG. Because of this variability, directly mapping neural signals onto decoding targets (e.g., text, speech, and music) is not optimal. Pre-trained foundation models can mitigate this by incorporating prior knowledge about the decoding targets, aligning them more closely with the neural signals.

To date, LLMs have been incorporated into BMI systems to enhance text decoding. A wide range of machine learning techniques  have been employed to increase the efficiency and accuracy of EEG-based spelling systems \citep{Speier16}. In practice, these language models can either auto-complete decoded words or be integrated into classifiers to refine the probability estimates of potential letters based on previously decoded ones. Leveraging language models has proven to significantly reduce word-error-rates, especially when decoding text from intracranial ECoG or Utah array during speech attempts \citep{Moses21,Metzger22,Metzger23,Willett21,willett2023high}. A notable recent study \citep{tang2023semantic} utilized a pre-trained GPT-2 model to interpret perceived speech from fMRI scans, converting neural patterns into text. This research, which involved over 16 hours of fMRI data from participants listening to stories, has showcased the potential of BMI in decoding imagined speech and even in cross-modal decoding, such as interpreting text representations of mental states during silent film viewing.

Foundation models have also been instrumental in enhancing the performance of BMI systems, especially in decoding audio and visual signals \citep{Metzger23,anumanchipalli2019speech,wang2020stimulus,wang2023distributed,takagi2023high,denk2023brain2music,bellier2023music,Meta23}. For instance, \cite{Metzger23} utilized a pre-trained speech generative model to decode clear speech from neural signals. Specifically, they used a sophisticated transformer-based speech encoding model (``HuBERT'') to learn a compact representation of speech, which was then transformed into high-quality speech using a pre-trained synthesizer. Beyond speech, music decoding has also seen progresses with the aid of generative AI. Multiple lines of recent research \citep{denk2023brain2music,bellier2023music} have demonstrated the feasibility of decoding music from neural signals using deep learning, with pre-trained models such as musicLM \citep{agostinelli2023musiclm}, to produce high-quality outputs. Similarly, image reconstruction from fMRI scans has achieved remarkable accuracy with the help of image generative models such as the VAE, GAN, and diffusion models \citep{takagi2023high,ferrante2023brain,VanRullen19_CB,huang2021deep,ozcelik2023natural}. 
In these studies, neural signals were first converted into latent representations, and then used to produce images through various generative models (Table~\ref{BMI}). For instance, a two-stage scene reconstruction framework called ``Brain-Diffuser" has been proposed: in the first stage, low-level image was first reconstructed via a very deep VAE, and in the second stage, a latent diffusion model conditioned on predicted multi-modal (text and visual) features was used to reconstruct high-quality images \citep{ozcelik2023natural}.

Remarkably, \cite{Meta23} developed an real-time visual decoding strategy from MEG recordings using a foundation model. The model consists of three modules: (i) pre-trained embedding obtained from images, (ii) an MEG module trained end-to-end, and (iii) a pre-trained image generator. Furthermore, the brain-to-image readout was decoded with  a foundational image model known as DINOv2.
The authors reported that MEG-based decoding can recover high-level visual features compared to fMRI-based decoding, offering a real-time BMI paradigm ($\sim$250 ms delay) for the human brain.

To date, most of brain decoding applications have been reported in human research since data format and acquisition  are relatively universal, which may not be the case in animal studies. Recently, built upon a foundation model known as Perceiver IO \citep{jaegle2022perceiver}, \cite{azabou2023unified} developed a new framework called POYO (Pre-training On manY NeurONs) for large-scale training transformer models end-to-end
on multi-session and across-individual electrophysiology datasets. POYO introduces  innovative spike-based tokenization strategies and used pre-trained models (with possible fine tuning) for neural population decoding; with a transformer architecture, POYO applies both cross-attention and self-attention  in the latent space after latent embeddings of neural events. Their work demonstrates that the power of transfer learning and transformer to achieve rapid and scalable neural decoding.  

\subsection{Data augmentation}

%Deep-learning models can process raw data, but first they must be trained with annotated information. 

 Machine learning-driven data augmentation techniques are beneficial to alleviate the sample imbalance or insufficiency problem \citep{Chawla02,HeGarcia09}. This is particularly important for improving the generalization ability of deep learning. Recently, data-centric deep learning or generative AI strategies (e.g., data regeneration and synthetic data generation) have been proposed to improve the consistency between the existing and augmented data, especially in clinical applications where labeled samples may be scarce  or the data privacy  is a concern \citep{zhang2022shifting}. For instance, 
 combining  RNN and GAN may help construct generative models of synthetic time series and impute missing sequences \citep{Yoon19,Lee21,Habashi23}. In one example, combined GAN and VAE models utilized three-dimensional convolution to model high-dimensional fMRI sensors with structured spatial correlations and the  synthesized datasets were then used to augment classifiers designed to predict cognitive and behavioral outcomes \citep{zhuang19}.  
 In another example,  an auxiliary classifier GAN (AC-GAN) was used to generate synthetic interictal epileptiform discharges (IED) from EEG recordings of epileptic patients \citep{Geng21,Geng21NER}. 

\cite{bird2021synthetic} employed an LLM (based on GPT-2) to augment the EEG/MEG dataset for a classification task. After initial training, the GPT model was used to generate realistic synthetic neural signal given corresponding classification labels as the augmented data;  a marginal improvement was reported in classification performance.

Recently, a text data augmentation approach based on ChatGPT (named AugGPT) \citep{dai2023auggpt}, has been developed to overcome the challenge of limited sample sizes in NLP tasks \citep{DA23}.  
 Specifically, sentences in the training set were rephrased into conceptually similar variations as the augmented data with the same label of the original sample.  The results showed that data augmentation based on such a large-scale pre-trained model increased the classification accuracy by a big margin in comparison with  standard data augmentation methods. However, more research is still needed to see whether similar techniques can apply to neural data augmentation.

\section{Discussion and conclusion }

\subsection{Crosstalk between AI and neuroscience}

 AI and neuroscience have been driving each other forward. Not only neuroscience has inspired the development of deep learning and AI technologies  \citep{hassabis2017neuroscience},  explainable AI and deep learning have also generated opportunities for in-depth neuroscience investigations \citep{richards2019deep,Saxe21}. For instance, biologically constrained CNN models have enabled neuroscientists to directly compare data in the visual cortex and uncover the underlying computational principles \citep{Yamins14,Yamins16,Shi22}.
Recently, \cite{Schneider23} proposed a contrastive learning-based neural network model for jointly modeling neural and behavioral dynamics. The SSL algorithm, known as CERBA, combining ideas from nonlinear independent component analysis (ICA) with contrastive learning, may identify  interpretable and consistent neural embeddings of high-dimensional neural recordings using auxiliary variables (such as time or behavioral measures). Importantly, it  can  generate embeddings across multiple subjects and cope with distribution shifts among experimental sessions, subjects, and recording modalities. In another example, \cite{Caucheteux23} applied deep language algorithms (based on GPT-2) to predict nearby words and discovered that the activations of language models linearly map onto the brain responses to speech, and these predictions are organized hierarchically in frontoparietal and temporal cortices. These findings illustrate the synergy between neuroscience and AI can largely improve our understanding of human cognition.

It is also worth mentioning that current AI technologies have relied on oversimplified models of neural systems. First and foremost, the standard artificial neurons in deep neural networks are ``point neurons" that focus on somatic computation, yet the importance of nonlinear dendritic computation has been ignored.
However, it has been known that the dendrite also plays an important role in neuronal computations and biological learning, such as enhancing expressivity of single neurons, improving neuronal resources and generalization abilities, utilizing internal learning signals, and enabling continual learning, contextual representation, and predictive coding \citep{ACHARYA22,Hodassman22,Hawkins16}. Deep learning models have the potential to reproduce computational complexity of biologically realistic neurons' I/O properties \citep{BENIAGUEV21}. Second, brain oscillations are important hallmarks in representing neural dynamics for a wide range of tasks in cognition, attention, memory, decision-making, and sensorimotor integration. Future development of next-generation neuroAI models and biologically plausible learning algorithms remains a central research direction to transform a ``black-box'' to ``glass box'' model while achieving  a good trade-off between performance and interpretability.

\subsection{Outlook and outstanding questions}

Looking ahead,  foundation models and generative AI will anticipate a rapid research growth in method development and applications, especially in brain imaging and large-scale neural and behavioral data analyses. 
In clinical applications, foundation models and generative AI may have a translational impact on personalized medicine. 
A growing number of ChatBots, such as ChatGPT and Bard, can play an active role in mitigating the worldwide crisis in mental health \citep{Chen22}. In multi-modal BMI systems, generative AI  will help combine speech, vision and motor modalities to improve the functionality and decoding accuracy. 
Future developments of brain-to-content neurotechnologies  may have promising applications in  immersive virtual reality, video games, marketing, and personalizied education.

Finally, we present several outstanding questions that might motivate future research in the intersection of AI and neuroscience. 
\begin{itemize}
\item 
Since the majority of  foundation models have been trained on single-modal data,  it is unclear whether  the model would benefit from training based on multi-modal or cross-modal data when the decoding domain is on single modality. For instance, in simultaneous EEG-fMRI recordings, can we train a foundation model based on their joint measurements, and then apply the pre-trained model in EEG-alone or fMRI-alone decoding analysis? 
While  the prior knowledge of the cross-modal relationship may be beneficial, the variability in SNR and spatiotemporal resolution between two modalities  may create practical barriers. Furthermore, it remains an open question how  we should apply SSL   to identify an optimal analysis pipeline for multi-modal neuroimaging data. 

\item Representation learning and foundation models have great potentials in RL, including end-to-end policy learning  \citep{bahl2020neural} and multi-agent communications \citep{Foerster16}. However, it remains unclear how well the foundation models and learned embedding representations can generalize across tasks in RL. For instance, RL algorithms have  been developed in BMI applications, enabling individuals with motor disabilities to control external devices using neural signals. It still needs to be thoroughly tested whether the pre-trained policy can generalize across subjects, tasks, and environments. Identifying common as well as individualized decision-making or control policy under the new representation learning paradigm will continue to be an active research topic.

\item  While ChatGPT can be used as an interface between users and external systems serving as a bridge between individuals with limited mobility and the external world, 
it is vital to revolutionize communication capabilities of BMIs by translating thoughts into text-based information and  refining the dynamics of human-machine interaction.
However, it remains unclear how    ChatGPT or GPT-like models can be optimally integrated into the BMI systems. 
Furthermore, can we adapt these models or generative AI to interpret and produce text that syncs flawlessly with a user's intentions while abiding by ethical and privacy mandates?
The recurrent engagement of users with ChatGPT offers prospects to transform the lives of those with disabilities
and to develop personalized and adaptable BMI systems, escalating user gratification and optimizing system outputs.

\item Ongoing research has continued producing new frontiers in foundation models and generative AI, such as the new autonomous AI agent tools (AutoGPT, MetaGPT and AutoGen) (see a compiled list at \url{https://github.com/steven2358/awesome-generative-ai}).
Integration of these emerging AI technologies into neuroscience applications  presents more challenges and opportunities.

\end{itemize}

In conclusion, many research areas in neuroscience have greatly benefited from BigData-empowered machine learning. Exploitation of large-scale foundation models, generative AI, and transfer learning tools will enable us to potentially probe neuroscience questions and brain-to-content technology in new dimensions. The landscape of neuroscience research is rapidly changing, and our imagination is only the limit for unlimited creativity. 
We hope this mini-review will inspire  more exciting work  in the near future.

\section*{Funding}

The work was partially supported by grants MH118928, DA056394, NS123928, NS121776, MH132642, and NS135170 from the US National Institutes of Health.

\section*{Declaration of Competing Interest}

The authors declare no competing interests.

\section*{Data availability}

No data was used for the research described in the current article.

\section*{Acknowledgments}

The authors thank Dr. Ryota Kobayashi and Dr. Ken Nakae for the invitation to participate in a Special Session at the Annual Japanese Neuroscience Meeting held in Sendai, Japan on August 3, 2023, which motivated the writing of this review.  

%% numbering style
 %\bibliographystyle{elsarticle-num} 
 %% Harvard style
\bibliographystyle{elsarticle-harv}
\bibliography{biblio}

\end{document}